\documentclass[aps,pra,twocolumn,showpacs,superscriptaddress,reprint]{revtex4}
\usepackage{graphicx}
\usepackage{amssymb}
\usepackage{mathrsfs}
\usepackage{amsmath}
\usepackage{bm,hyperref}
\usepackage{subfigure}
\allowdisplaybreaks[4]
\begin{document}
\title{Coherent-State Approach for Majorana representation}
\author{H. D. Liu}
\affiliation{Beijing Computational Science Research Center, Beijing 100084, China}
\affiliation{Center of Quantum Sciences and School of Physics, Northeast Normal University, Changchun 130024, China}

\author{L. B. Fu}
\affiliation{National Laboratory of Science and Technology on Computational Physics,
Institute of Applied Physics and Computational Mathematics, Beijing 100088, China}
\author{Xiaoguang Wang\footnote{xgwang@zimp.zju.edu.cn}}
\affiliation{Beijing Computational Science Research Center, Beijing 100084, China}
\affiliation{Zhejiang Institute of Modern Physics, Department of Physics, Zhejiang University, Hangzhou 310027, China}

\date{\today}
\begin{abstract}
By representing a quantum state and its evolution with the majorana stars on the Bloch sphere, the Majorana representation (MR) provide us an intuitive way to study a physical system with SU(2) symmetry. In this work, based on coherent states, we propose a method to establish generalization of MR for a general symmetry. By choosing a generalized coherent state as a reference state, we give a more general MR for both finite and infinite systems and the corresponding star equations are given. Using this method, we study the squeezed vacuum states for three different symmetries, Heisenberg-Weyl, SU(2) and SU(1,1), and express the effect of squeezing parameter on the distribution of stars. Furthermore,  we also study the dynamical evolution of stars for an initial coherent state driven by a nonlinear Hamiltonian, and find that at a special time point, the stars are distributed on two orthogonal large circles.
\end{abstract}
\maketitle
\section{Introduction}
The Majorana representation (MR), which provides an intuitive picture to study a physical system with a high dimensional projective Hilbert space\cite{Majorana1932}, has attracted revived attention in recent years. Despite being introduced about 80 years ago, this representation, which endows quantum state with visualization, has become an efficient tool to study the symmetric related feature of quantum system, such as spinor boson gases\cite{Kawaguchia2012,Stamper-Kurn2013,Lian2012,Cui2012,Barnett2009,Lamacraft2010}, multilevel qubits\cite{UshaDevi2012}, and Lipkin-Meshkov-Glick model\cite{Ribeiro2007,Ribeiro2008}, since it naturally provides an intuitive way to study the geometrical perspectives of these systems, e.g., geometric phase\cite{Bruno2012,Liu2014,Tamate2011,Ogawa2015,Yang2015}, entanglement\cite{Bastin2009,Ribeiro2011,Markham2010,Ganczarek2012,Mandilara2014,Baguette2015}.

As we know, a two-level pure state can be described by a point on the Bloch sphere, and its evolution is perfectly represented by the trajectory of the point on the sphere. However, it seems hard to extend this geometric interpretation to a higher dimensional quantum states, since it is difficult to visualize the trace in high dimensional space. Faced with this, the early ingenious work of Majorana told us that we can study the problem from a different perspective: including more points on the two-dimensional Bloch sphere instead picturing one single point on a high dimensional geometric structure. In MR, one can describe a spin-$J$ state intuitively by $2J$ points on the Bloch sphere. These points are called Majorana stars of the state. Therefore, the physics information of a spin state, such as dynamic evolution, geometric phase, mutiparticle entanglement, can be represented by these stars.

The fly in the ointment is that this elegant geometric representation can only be used to study a pure spin state which has SU(2) symmetry. With the increasing attention of MR, how to extend the this representations to mixed states or pure states with other symmetires becomes a fascinating problem. Recently, Giraud et al. propose an generalization to arbitrary spin-$j$ mixed state of the MR in terms of tensors that share the most
important properties of Bloch vectors based on covariant matrices introduced by
Weinberg\cite{Giraud2015}. Moreover, for any $n$-dimensional pure state, the parameterization process in Majorana representation can also be borrowed to define $n-1$ stars on the Bloch sphere, but the symmetry features carried by the state might not be properly presented. It is a natural question to ask whether a similar geometric representation can be found for a general pure quantum state without loss of symmetry information. Inspired by the relation between Majorana representation and Schwinger Boson, we find the answer lies with the generators of the SU(2). For an particular symmetry described by a Lie group, similar with SU(2), its generators can always correspond to a set of ladder and number operators, which can determine a parameterize way of the state as in SU(2).

In this work, we present a new extension for the Majorana representation. Based on the definitions of generalized coherent states\cite{Peremolov1986} with the ladder and number operators, we establish a procedure of establishing representation like MR for an arbitrary symmetry and chose the coherent state as a reference. We show that this coherent state approach for Majorana representation can not only be used to typical states like coherent states and squeezed vacuum states for a particular symmetry, such as Heidelberg-Weyl (HW), SU(2), and SU(1,1) symmetries, but also can reproduce the change of symmetry in the period evolution of a quantum state. In this respect, it provides an intuitive way to study the quantum system which carries a particular symmetry.

This paper is organized as follows. In Sec. II A, we introduce the Majorana represent (MR) and the coherent state in MR. Then, by using the coherent state and ladder operators, we establish a coherent state approach for an arbitrary symmetry and obtain a new representation of their states by stars on the Bloch sphere in Sec. II B. In Sec. III  this
coherent state approach representation is applied to three particular symmetries of Heidelberg-Weyl, SU(2), SU(1,1), and obtain three star equations, respectively. In Sec. IV, we investigate the squeezed vacuum states in these three symmetries by the stars in the coherent state approach representation. In Sec V, a nonlinear system is studied to illustrate our theory. A brief discussion and summary are given in Sec. VI.

\section{Majorana representation and its coherent-state approach}
\subsection{Majorana representation }
We first introduce the MR which was developed for spins~\cite{Majorana1932}. A generic spin-$j$ state
\begin{equation}
\begin{aligned}
|\psi\rangle^{(j)}&=\sum_{m=-j}^{j}C_m|j,m\rangle\\
&=\sum_{n=0}^{2j}C_{n-j}|n\rangle_j\\
&=\sum_{k=0}^{2j}C_{j-k}|2j-k\rangle_j,
\end{aligned}\label{gstate}
\end{equation}
where $|n\rangle_j\equiv |j,-j+n\rangle$, $k=2j-n$, and $n, k$ are integers.
It is instructive to write the above state under the two-boson representation. Formally,
 the spin basis state $|j,m\rangle$ corresponds to a two mode boson state $|j+m,j-m\rangle$.
Consequently, in the form of boson creation operators $\hat{a}^\dagger$ and $\hat{b}^\dagger$, the spin state $|\psi\rangle^{(j)}$ can be written as
\begin{equation}
\begin{aligned}
|\psi \rangle^{(j)}
&=\sum_{m=-j}^{j}\frac{C_{m}{a}^{\dag (j+m)}{b}%
^{\dag (j-m)}}{\sqrt{(j+m)!(j-m)!}}%
|00\rangle
\\
&=\sum_{n=0}^{2j}\frac{C_{n-j}{a}^{\dag n}{b}%
^{\dag (2j-n)}}{\sqrt{n!(2j-n)!}}%
|00\rangle\\
&=\sum_{k=0}^{2j}\frac{C_{j-k}{a}^{\dag (2j-k)}{b}%
^{\dag k}}{\sqrt{k!(2j-k)!}}%
|00\rangle.
\end{aligned}
\label{MSRP}
\end{equation}

Then, we meet a homogeneous polynomial of degree $2j$
\begin{equation}
f(x,y)=\sum_{k=0}^{2j}\frac{C_{j-k}{x}^{(2j-k)}{y}%
^{k}}{\sqrt{k!(2j-k)!}}.
\end{equation}
This can be further written as the following form
\begin{equation}
f(x,y)=(-y)^{2j}\sum_{k=0}^{2j}\frac{(-1)^kC_{j-k}{z}^{(2j-k)}}{\sqrt{k!(2j-k)!}},
\label{fff}
\end{equation}
where $z=-x/y$. Then, by solving the following star equation
\begin{equation}
\sum^{2j}_{k=0}\frac{(-1)^kC_{j-k}}{\sqrt{(2j-k)!\,k!}}z^{2j-k}=0,\label{mreq}
\end{equation}
we may find $2j$ roots $z_1,z_2,\cdots,z_{n}$. Finally, the polynomial (\ref{fff})
can be written as a factorized form
\begin{equation}
f(x,y)=(-y)^{2j}\prod_{k=1}^{2j} (z-z_k)=\prod_{k=1}^{2j} (x+z_k y).
\label{ffff}
\end{equation}

Using Eq.~(\ref{ffff}), the state $|\psi \rangle^{(j)}$ (\ref{MSRP}) becomes
\begin{equation}
|\psi \rangle^{(j)}=\prod_{k=1}^{2j} (a^\dagger+z_k b^\dagger)|00\rangle.\label{state}
\end{equation}
There are $2j$ complex numbers $z_k$ determined by Eq.~(\ref{mreq}). These numbers completely
describe the state and can be geometrically described by $2j $ points on a plane or on a unit sphere via relation
\begin{equation}
z_k=\tan\frac{\theta_k}{2}e^{i\phi_k}, \theta_k\in [0,\pi], \phi_k\in [0,2\pi],\label{star}
\end{equation}
where $\theta_k$ and $\phi_k$ are the spherical coordinates.
Therefore, any spin state $|\psi \rangle^{(j)}$ and its
evolution can be depicted by these points which are called Majorana stars.
Substituting Eq.~(\ref{star}) into (\ref{state}) leads to
\begin{equation}
\begin{aligned}
|\psi \rangle^{(j)}&=\left(\frac{1}{\cos\theta_k}\right)^{2j}\prod_{k=1}^{2j} (\cos\theta_k a^\dagger+\sin\theta_k e^{i\phi_k}b^\dagger)|00\rangle.\\\label{state2}
&=\left(\frac{1}{\cos\theta_k}\right)^{2j}\prod_{k=1}^{2j}c^\dagger_k |00\rangle,
\end{aligned}
\end{equation}
where
\begin{equation}
c^\dagger_k=\cos\theta_k a^\dagger+\sin\theta_k e^{i\phi_k}b^\dagger
\end{equation}
are also bosonic creation operators. This equation is another form of the spin state.

Now, as a example,  we consider a spin coherent state (SCS) defined by~\cite{SCS1,SCS2}
\begin{equation}
|\eta\rangle_j=\left(1+|\eta|^2\right)^{-j}\sum_{n=0}^{2j}\begin{pmatrix}
                                                          2j \\
                                                          n \\
                                                        \end{pmatrix}^{1/2}\eta^n |n\rangle_j,
                                                        \label{coherent}
\end{equation}
where $\eta$ is a  complex number. Comparing Eq.~(\ref{gstate}) and (\ref{coherent}), one finds
\begin{equation}
C_{j-k}=\left(1+|\eta|^2\right)^{-j}\begin{pmatrix}
                                                          2j \\
                                                          k \\
                                                        \end{pmatrix}^{1/2}\eta^{2j-k} ,
                                                        \label{coherent1}
\end{equation}
For finding the majorana stars, substituting the above equation into Eq.~(\ref{mreq}) leads to
\begin{equation}
(-1)^{2j}(1+|\eta|^2)^{-j}\sqrt{(2j)!}(1-\eta z)^{2j}=0.
\end{equation}
Thus, there are $2j$-fold roots $z=\eta^{-1}$ and  $2j$ stars coincide in one single point on the Bloch sphere.
There are still one point even in the case of $j\rightarrow\infty$\cite{Penrose1990}. So, we can choose the coherent state as a reference state when we
intend to generalize the MR to more general systems including other finite or infinite systems. Next, based on the coherent state defined on a Lie group, we use the coherent-state
approach to define new MR.

\subsection{Coherent-state approach}

For the system with the symmetry of a Lie group $G$, there exists a method to construct the coherent states \cite{Peremolov1986}.
The generators $L_i$ of group $G$ satisfying the commutation relation
\begin{equation}
[L_i,L_j]=C_{ijk}L_k
\end{equation}
with structure constants $C_{ijk}$.
 We may construct the ladder operators ${A}$ and ${A}^\dag$ by the linear combination of $\{{L}_j\}$ and define number operator ${\cal N}$ via a certain $ L_\lambda$ which satisfy
\begin{equation}
\begin{aligned}
{A}|m\rangle&=\sqrt{\varepsilon_m}|m-1\rangle,~{A}^\dagger|m\rangle=\sqrt{\varepsilon_{m+1}}|m+1\rangle,\\
{\cal N}|m\rangle&=m|m\rangle,
\end{aligned}
\label{ladder}
\end{equation}
where $\varepsilon_1,\varepsilon_2,\dots,\varepsilon_m$ is a sequence of
positive numbers.

Using the creation operator, a generic state $|\psi\rangle=\sum_mC_m|m\rangle$ can be factorized as
\begin{equation}
|\psi\rangle=\sum_m\frac{C_m{A}^{\dag m}}{\sqrt{\epsilon_m!}}|0\rangle
\sim\prod_m({A}^\dag+\lambda_m)\rangle|0\rangle.
\label{psiA}
\end{equation}
where $\epsilon_m!=\epsilon_1\epsilon_2\cdots\epsilon_m$ and the roots are determined via
\begin{equation}
\sum_m\frac{C_mz^m}{\sqrt{\epsilon_m!}}=0.
\end{equation}

Interestingly, if we decompose the complex numbers $\lambda_m$ as $\lambda_m=\tan\frac{\theta_m}{2}e^{i\phi_m}$, we can represent state $|\psi\rangle$ as points $(\theta_m,\phi_m)$ on the Bloch sphere as the stars in MR. However, the choice of ladder operators in Eq.(\ref{ladder}) is not unique. Since, we can add any coefficient which is the function of $m$, the new sequence of $\epsilon'_m$ will still hold Eq.(\ref{ladder}). This uncertainty of ladder operator seemingly become an obstacle to establish a symmetric-related representation. At this moment, we look back at the last subsection, the answer emerges naturally. The interesting representation that all the stars of the coherent state in MR accumulating on one point on the Bloch sphere provides us a natural reference to fix the choice of the ladder operator $ {A}^\dag$. Therefore, we call this new representation as the coherent state approach (CSA) for MR.

The coherent state can be defined as\cite{Peremolov1986}
\begin{equation}
|\psi_C\rangle=D(\bm{\tau})|\psi_0\rangle,
\label{CSD}
\end{equation}
where $|\psi_0\rangle=|0\rangle$ is a fixed state which can be chosen as the eigenstate of some generator $\hat L_\lambda$  and the symmetry-related operator $D(\bm{\tau})$ is constructed by all the ladder operators $A$ and $A^\dag$. Since $D(\bm{\tau})$ is an unitary operator, it can be written as
\begin{equation}
D(\bm{\tau})=\exp\left(-i\sum^n_{j\neq\lambda}\tau_j L_j\right)\sim\exp\left(\alpha A^\dag-\alpha^*A\right).
\label{D}
\end{equation}
Note that, ladder operators $A$ and $A^\dag$ are constructed by $\{{L}_j\}$. Therefore, we may have several pairs of ladder operators. For simplicity, we only consider one pair of ladder operators in this paper.

Technically, we can establish this new representation in three steps:

(i) Constructing the ladder operator $A^\dagger$ and the coherent state $|\alpha\rangle$ of the system with a particular symmetry. The operator $A^\dagger$ can be constructed by generators of the corresponding Lie group\cite{Peremolov1986}.
Suppose using the commutation relation between the ladder operators $A$ and $A^\dag$, the coherent state can be defined as
\begin{equation}
|\alpha\rangle\sim e^{\alpha  A^{\dagger}}|0\rangle
\label{CS}
\end{equation}

(ii) Using the coherent state as a reference to fix the ladder operator and majorana points. To represent the coherent state as one point on the Bloch sphere, we define a nonlinear creation operator $ {\tilde{A}}^\dagger=f(  N)  A^\dagger$ to change the form of Eq. (\ref{CS}) into
\begin{equation}
|\alpha\rangle\sim\sum^{n}_{l=0}\frac{\alpha^l {\tilde{A}}^{\dagger l}}{l!(n-l)!)}|0\rangle\sim( {\tilde{A}}^{\dagger}+\alpha^{-1})^n|0\rangle.
\end{equation}
Correspondingly, Eq. (\ref{psiA}) becomes
\begin{equation}
|\psi\rangle=\frac{1}{N}\prod_{m=1}^n( {\tilde{A}}^\dagger+\lambda_m)|0\rangle.
\label{psiAt}
\end{equation}
Define the new complex coefficients as $\lambda_m=\tan\frac{\theta_m}{2}e^{i\phi_m}$, we have $n$  Majorana stars on the unit sphere. With this choice, one can guarantee that
the stars for the coherent state coincide on one point, just like the case of spins.

(iii) Establishing the equation for  Majorana stars. If we meet an infinite system, the cutoff can be made as
the all excitations for a physical state cannot be infinite.
With this procedure, we next apply this CSA to some physical systems with particular symmetries.

\section{Applications for several symmetries}
\subsection{Spin state for SU(2) symmetry}
First, we need to guarantee our CSA can reproduce the MR for SU(2) symmetry.
The spin operators ${ J_x,   J_y,   J_z}$ as the generators of SU(2) have the commutation relations
\begin{equation}
[  J_z,  J_\pm]=\pm  J_\pm,~~[  J_+,   J_-]=2  J_z,
\end{equation}
where $  J_\pm=  J_x\pm i  J_y$ are ladder operators.  We have the following relations
\begin{equation}
\begin{aligned}
  J_+|n\rangle_j&=\sqrt{(n+1)(2j-n)}|n+1\rangle_j,\\
  {\cal N}_j|n\rangle_j&=(J_z+j)|n\rangle_j=n|n\rangle_j.\label{up}
\end{aligned}
\end{equation}
Then, we arrive at
\begin{equation}
|n\rangle_j=\sqrt{\frac{(2j-n)!}{n!(2j)!}}J^n_+|0\rangle_j,\label{upup}
\end{equation}
The coherent state for spins is well-defined and the related displacement operator $D_j({\tau})$ can be defined as
\begin{equation}
D_j(\tau)=e^{\tau  J_+-\tau^*  J_-}=e^{\eta  J_+}e^{\bar{\eta}  J_z}e^{-\eta*  J_-}
\end{equation}
with $\eta=\tan|\tau|e^{i\mathrm{arg}(\tau)}$ and $\bar{\eta}=\ln(1+|\tau|^2)$. The SCS is defined as
\begin{equation}
|\eta\rangle_j=D_j(\tau)|0\rangle_j\sim e^{\eta  J_+}|0\rangle_j.
\label{SCS2}
\end{equation}

From Eq.~(\ref{upup}), the general form of the spin state $|\psi\rangle^{(j)}$ (\ref{gstate}) becomes
\begin{equation}
|\psi\rangle^{(j)}=\sum_{n=0}^{2j}C_n|n\rangle_j=\sum_{n=0}^{2j}\sqrt{\frac{(2j-n)!}{n!(2j)!}}C_n  J^n_+|0\rangle_j.\label{ggstate}
\end{equation}
One may define the stars from the above form via the following star equation
\begin{equation}
\sum_{n=0}^{2j}\sqrt{\frac{(2j-n)!}{n!(2j)!}}(-1)^nC_n  z^n=0.
\end{equation}
However, this equation is different from the star equation for spins (\ref{mreq}). In other words, if we solve this equation
for SCS, there will be $2j$ different stars and cannot guarantee all stars coincide at a single point.

We solve this puzzle by introducing the nonlinear creation operator as
\begin{equation}
{\tilde{J}}_+= f({\cal N}_j){J}_+= \frac{1}{2j- {\cal N}_j+1}{J}_+,
\label{superj}
\end{equation}
which have the property
\begin{equation}
{\tilde{J}}_{+}^n=\left(\prod_{k=0}^{n-1}f({\cal N}_j-k)\right) {{J}}_{+}^n.
\end{equation}
Using the above equation and Eq.~(\ref{upup}), let ${\tilde{J}}_{+}^n$ act on state $|0\rangle_j$, we obtain
\begin{equation}
|n\rangle_j=\sqrt{\frac{(2j)!}{n!(2j-n)!}}\tilde{J}^n_+|0\rangle_j,\label{upupup}
\end{equation}
Then, the general state can be written as
\begin{equation}
|\psi\rangle^{(j)}=\sum_{n=0}^{2j}\begin{pmatrix}
                                                          2j \\
                                                          n \\
                                                        \end{pmatrix}^{1/2}C_n (-1)^n  (-\tilde{J}_+)^n |0\rangle_j.\label{gggstate}
\end{equation}
From the above equation, finally, we obtain the star equation by considering $-\tilde{J}_+$ as a number
\begin{equation}
\sum_{n=0}^{2j}\begin{pmatrix}
                                                          2j \\
                                                          n \\
                                                        \end{pmatrix}^{1/2}(-1)^n C_n z^n=0.
\label{su2star}
\end{equation}
This equation is a little different but essentially have the same roots of the star equation (\ref{mreq}). Obviously, from this form, all stars for the SCS (\ref{coherent}) coincides.
Next, using this coherent-state approach, we generalize MR from finite SU(2) systems to systems with infinite dimensions.

\subsection{Single Mode Boson state for HW symmetry}
In a similar way with the above discussion, we consider the bosonic single-mode system which has HW symmetry.
Its generators are the boson creation operator $  a^\dagger$ and annihilation operator $  a$ and the unity operator $I$,
which satisfy the commutation relations
\begin{equation}
[  a,  a^\dag]=  I, [  a,   I]=[  a^\dag,   I]=0.
\end{equation}
The ladder operator $a^\dagger$ and bosonic number operator $ {\cal N}=  a^\dagger  a$ for the Fock basis $\{|m\rangle\}$ satisfy
\begin{equation}
  a^\dagger|n\rangle=\sqrt{n+1}|n+1\rangle, ~~~~~  {\cal N}|n\rangle=n|n\rangle.
\end{equation}
Thus, a single mode boson state takes the form
\begin{equation}
|\psi\rangle^{}=\sum_{n=0}^{\infty}C_n|n\rangle=\sum_{n=0}^{\infty}\frac{C_n  a^{\dagger n}}{\sqrt{n!}}|0\rangle.\label{ggg}
\end{equation}

The coherent state can be obtained by action of the displacement operator,
\begin{equation}
D_a(\alpha)=e^{\alpha  a^\dag-\alpha^*a}=e^{\alpha  a^\dag}e^{-|\alpha|^2/2}e^{-\alpha^*  a}
\end{equation}
on the vacuum state. Then, the coherent state is given by
\begin{equation}
|\alpha_{a}\rangle=e^{-|\alpha|^2/2}e^{\alpha a^\dagger}|0\rangle=e^{-|\alpha|^2/2}\sum_{n=0}^\infty \frac{\alpha^n}{\sqrt{n!}}|n\rangle.
\label{HWC}
\end{equation}

Different with the situation of SU(2), if we want to establish a geometric representation by stars, we need to truncate the infinite to a finite number $N_c$.
Define a nonlinear creation operator as
\begin{equation}
{\tilde{a}}^\dagger= \frac{1}{N_c- {\cal N}+1}{a}^\dagger,
\label{supera}
\end{equation}
which obeys
\begin{equation}
\tilde{a}^{\dagger n}=\left(\prod_{m=0}^{n-1}\frac{1}{N_c-{\cal N}+m+1}\right)a^{\dagger n}.
\end{equation}
Acting on the vacuum state leads to
\begin{equation}
\tilde{a}^{\dagger n}|0\rangle=\frac{(N_c-n)!\sqrt{n!}}{N_c!}|n\rangle.
\end{equation}
Thus, we obtain
\begin{equation}
|n\rangle=\frac{N_c!}{(N_c-n)!\sqrt{n!}}\tilde{a}^{\dagger n}|0\rangle.
\label{expap}
\end{equation}
Finally, we can write the general state (\ref{ggg}) in terms of $-\tilde{a}^\dagger$ as
\begin{equation}
|\psi\rangle=\sum_{n=0}^{N_c}\frac{N_c!}{(N_c-n)!\sqrt{n!}}(-1)^nC_n(-\tilde{a}^{\dagger})^{n}|0\rangle
\end{equation}
The star equation for the boson is given by
\begin{equation}
\sum_{n=0}^{N_c}\frac{N_c!}{(N_c-n)!\sqrt{n!}}(-1)^nC_nz^{n}=0.
\label{HWstar}
\end{equation}
This star equation is applicable to all pure states of a boson. Obviously, for the coherent state, all the stars coincide.

\subsection{State for SU(1,1) symmetry}
Another useful symmetry is SU(1,1) symmetry, which has been widely applied to study spin squeezing in quantum metrology\cite{Ma2011,Berrada2013} and some nonconservative physical systems\cite{Gerry1987,Choi2009,Wu2007,Song2003}. Similar with group SU(2), group SU(1,1) also has three generators $  K_1,   K_2$ and $  K_3$ which satisfy
\begin{equation}
[  K_0,  K_\pm]=\pm   K_\pm,~~~[  K_-,  K_+]=2  K_0.
\end{equation}
The irreducible representation is
\begin{equation}
\begin{aligned}
  K_+|k,n\rangle&=\sqrt{(n+1)(2k+n)}|k,n+1\rangle,\\
  K_0|k,n\rangle&=(k+n)|k,n\rangle\label{kkk}
\end{aligned}
\end{equation}
with $  K_\pm=(K_1\pm iK_2)$.
It is easy to verify that the quadratic operator
\begin{equation}
 {C}_2=  K_0^2-  K^2_1-  K^2_2=k(k-1)
\end{equation}
is invariant (the Casimir operator) with real number $k$ (Bargmann index).
Basis vectors $|k,m\rangle$ marked by an integer $m$ are the eigenvectors of the operator $K_0$. So, one can define a number operator ${\cal N}_k=K_0-k$,
which satisfy
\begin{equation}
{\cal N}_k |n\rangle_k=n|n\rangle_k,
\end{equation}
where $|n\rangle_k\equiv |k,n\rangle$.

From Eq.~(\ref{kkk}), we have
\begin{equation}
|n\rangle_k=\sqrt{\frac{\Gamma (2k)}{n!\Gamma
(2k+n)}}  K_+^n|0\rangle_k.
\end{equation}
Therefore, the general state $|\psi\rangle^{(k)}=\sum_{n=0}^\infty C_n|n\rangle_k$ takes the form
\begin{equation}
|\psi\rangle^{(k)}=\sum_{n=0}^\infty\sqrt{\frac{\Gamma (2k)}{n!\Gamma
(2k+n)}}C_n  K^n_+|0\rangle_k.
\end{equation}
The displacement operator for SU(1,1) system is defined by
\begin{equation}
D_k(\tau)=e^{\tau  K_+-\tau^*  K_-)}=e^{\beta  K_+}e^{\eta  K_Z}e^{\beta^*  K_-}
\end{equation}
with $\beta=\tanh|\tau|e^{i\mathrm{arg}(\tau)}$ and $\eta=-\ln(1-|\tau|^2)$.
The coherent state is defined as\cite{Peremolov1986}
\begin{equation}
|\beta\rangle_k\sim e^{\beta K_+}|0\rangle_k=\sum_{n=0}^\infty \sqrt{\frac{\Gamma(2k+n)}{n!\Gamma(2k)}}\beta^n|n\rangle_k.
\label{KCS}
\end{equation}

Again we truncate the upper to a finite number $N_c$ and define the nonlinear operator
\begin{equation}
{\tilde{K}}_+= \frac{1}{N_c-{\cal N}_k+1}{K}_+,
\label{superk}
\end{equation}
which obeys
\begin{equation}
\tilde{K}_+^{n}=\left(\prod_{m=0}^{n-1}\frac{1}{N_c-{\cal N}_k+m+1}\right)K_+^{n}.
\end{equation}
Acting on the vacuum state leads to
\begin{equation}
\tilde{K}_+^{n}|0\rangle=\frac{(N_c-n)!}{N_c!}\sqrt{\frac{n!\Gamma(2k+n)}{\Gamma(2k)}}|n\rangle_k.
\end{equation}
Thus, we obtain
\begin{equation}
|n\rangle_k=\frac{N_c!}{(N_c-n)!}\sqrt{\frac{\Gamma(2k)}{n!\Gamma(2k+n)}}\tilde{K}_+^{n}|0\rangle_k.
\end{equation}
From the above equation, the general SU(1,1) pure state is written as
\begin{equation}
|\psi\rangle^{(k)}=\sum_{n=0}^{N_c}C_n(-1)^n\frac{N_c!}{(N_c-n)!}\sqrt{\frac{\Gamma(2k)}{n!\Gamma(2k+n)}}(-\tilde{K})_+^{n}|0\rangle_k.
\end{equation}
The third star equation is then obtained as
\begin{equation}
\sum_{n=0}^{N_c}C_n(-1)^n\frac{N_c!}{(N_c-n)!}\sqrt{\frac{\Gamma(2k)}{n!\Gamma(2k+n)}}z^{n}=0.
\label{su11star}
\end{equation}
Substituting the coefficients of the coherent state (\ref{KCS}) into the above equation, one find again
all the stars coincide. Thus, we have obtained three star equations, respectively for SU(2), HW, and SU(1,1) systems.
Next, we apply these equations to real quantum states.

\section{Squeezed vaccum state in CSA}
So far, we have presented how to establish the CSA for some kinds of symmetries. This coherent-state based method provide us a geometric tool to study properties of quantum states. We now consider a class of quantum state, i.e., squeezed states.
First, we consider the typical single-mode squeezed vacuum (SMSV) state
\begin{equation}
\begin{aligned}
|\xi_a\rangle&=(1-|\xi_a|^2)^{1/4}e^{\frac{\xi}{2} {a}^{\dag2}}|0\rangle\\
&=(1-|\xi_a|^2)^{1/4}\sum _{n=0}^{\infty}\xi^{n}{\frac {\sqrt {(2n)!}}{2^{n}n!}}|2n\rangle,\\
\end{aligned}
\end{equation}
where $\xi_a$ is a complex number which satisfies $|\xi|<1$. So, the coefficients $C_{2n}$ is given by
\begin{equation}
C_{2n}=(1-|\xi_a|^2)^{1/4}\xi^{n}{\frac {\sqrt {(2n)!}}{2^{n}n!}},
\end{equation}
and $C_{2n+1}=0$. Substituting the above equation into the star equation for bosonic system (\ref{HWstar}), we arrive at
the star equation for the SMSV
\begin{equation}
\sum^{[N_c/2]}_{n=0}\frac{\xi^nz^{2n}}{(N_c-2n)!n!2^n}=\sum^{[N_c/2]}_{n=0}\frac{(-1)^n\tilde{z}^{n}}{(N_c-2n)!n!}=0,
\end{equation}
where $\tilde{z}=-z^2\xi/2$. By solving the star equation, one finds $[N_c/2]$ positive real roots $\tilde{z}_k$ (which cam be proved by using Descartes' rule of signs) and then we have
$N_c$ (even) roots given by
\begin{equation}
z_k=\pm i\sqrt{{2\tilde{z}_k}/{\xi}}.
\end{equation}
So, the roots appears in a pairwise way with phases $3\pi/2-\mathrm{Arg}(\xi)/2$ and $\pi/2-\mathrm{Arg}(\xi)/2$ because only even states are involved. For odd $N_c$, there exists an extra root $z_k=\infty$ corresponding to a star located on the south pole.

Similarly, one spin squeezed state for SU(2) symmetry is defined as\cite{Frahm1989,Nakajima1997}
\begin{equation}
\begin{aligned}
|\xi_J\rangle&\sim e^{\frac{\xi_j}{4j} {j}_+^{2}}|0\rangle\\
&\sim \sum _{n=0}^{[j]}\left(\frac{\xi_j}{4j}\right)^{n}\frac{\sqrt {(2n)!(2j)!}}{n!\sqrt{(2j-2n)!}}|2n\rangle.\\
\end{aligned}\label{su2coef}
\end{equation}
One can also define the squeezed state for SU(1,1) system as
\begin{equation}
\begin{aligned}
|\xi_k\rangle&\sim e^{\frac{\xi_k}{2} {K}_+^{2}}|0\rangle\\
&\sim\sum _{n=0}^{\infty}\left(\frac{\xi_k}{2}\right)^{n}\frac{\sqrt {(2n)!\Gamma(2k+2n)}}{n!\sqrt{\Gamma(2k)}}|2n\rangle\\
\end{aligned}
\label{su11coef}
\end{equation}
Substituting the coefficients in Eqs.~(\ref{su2coef}) and (\ref{su11coef}) into Eqs.~(\ref{su2star}) and (\ref{su11star}), respectively, we obtain
two star equations for the two squeezed states as
\begin{equation}
\begin{aligned}
&\sum^{[j]}_{n=0}\frac{(-1)^n\tilde{x}^n}{(2j-2n)!n!}=0,\\
&\sum^{[N_c/2]}_{n=0}\frac{(-1)^n\tilde{y}^n}{(N_c-2n)!n!}=0,
\label{svequations}
\end{aligned}
\end{equation}
where $\tilde{x}=-\xi_jz^2/4j, \tilde{y}=-\xi_kz^{2}/2.$  Thus, we see that for the three squeezed states, the three star equations are essentially identical. If we choose
other squeezed states, the star equations are different.

\begin{figure}[t]
\centering
\subfigure[]{\includegraphics[width=0.245\columnwidth]{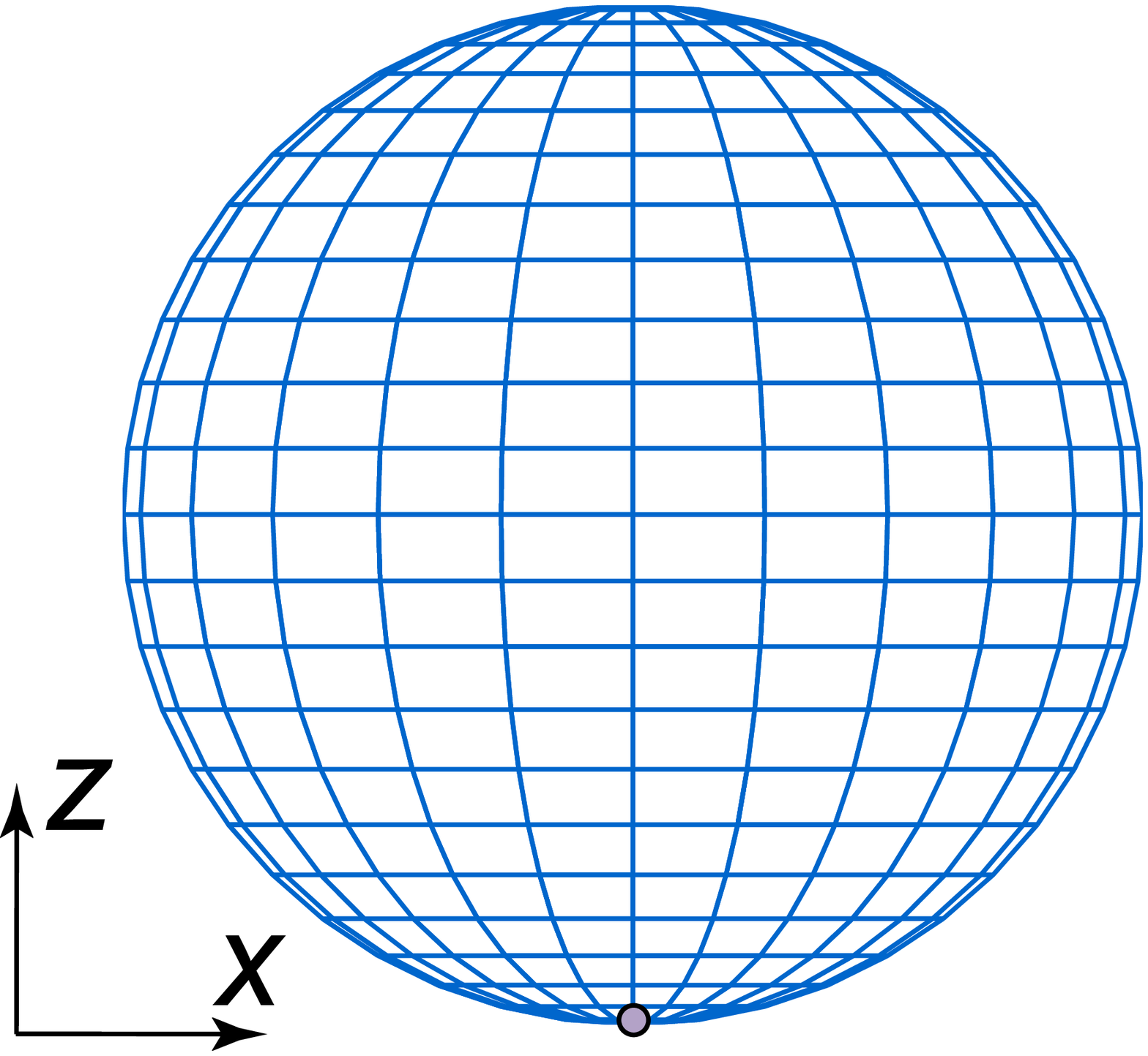}}
\subfigure[]{\includegraphics[width=0.22\columnwidth]{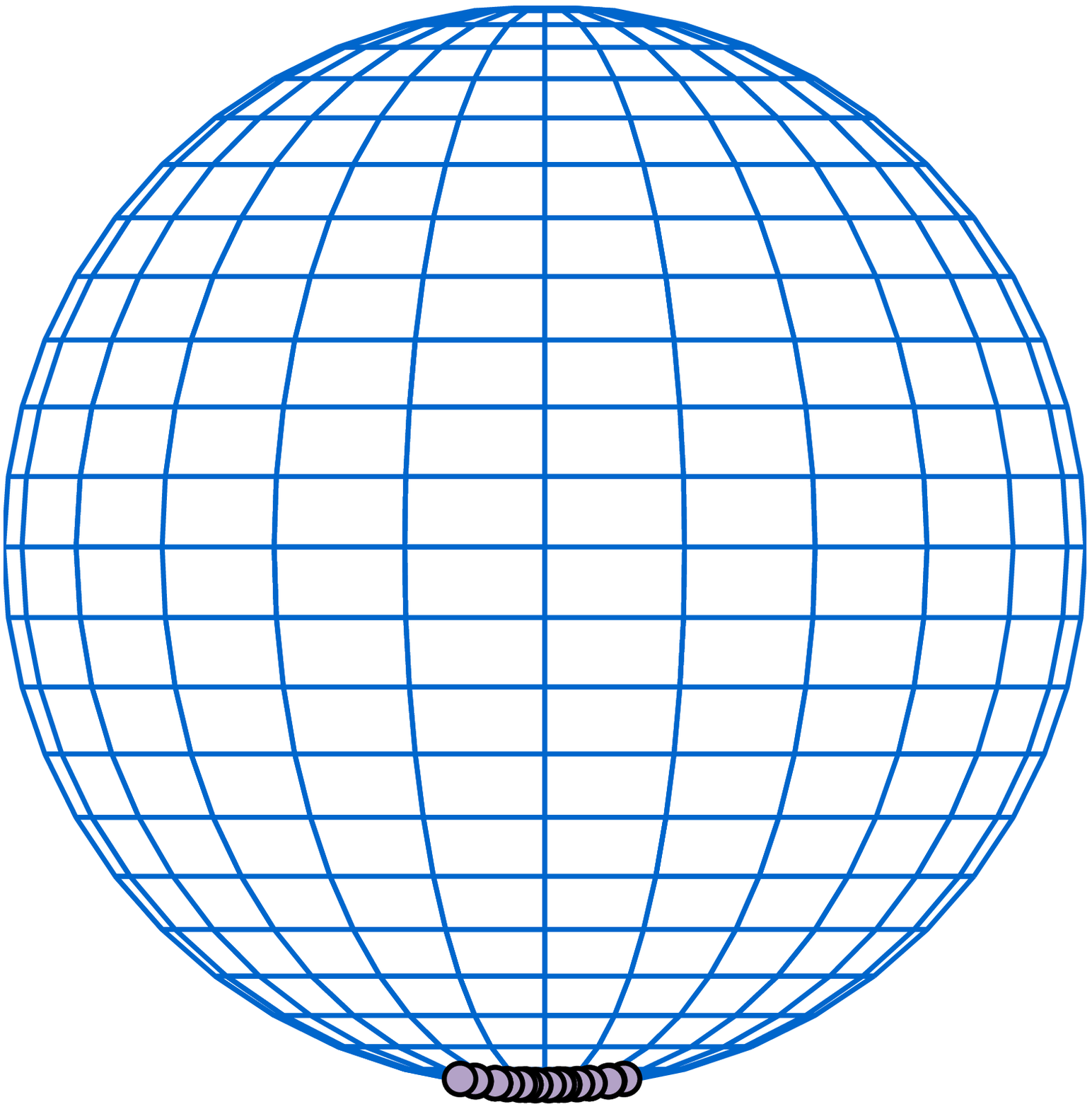}}
\subfigure[]{\includegraphics[width=0.22\columnwidth]{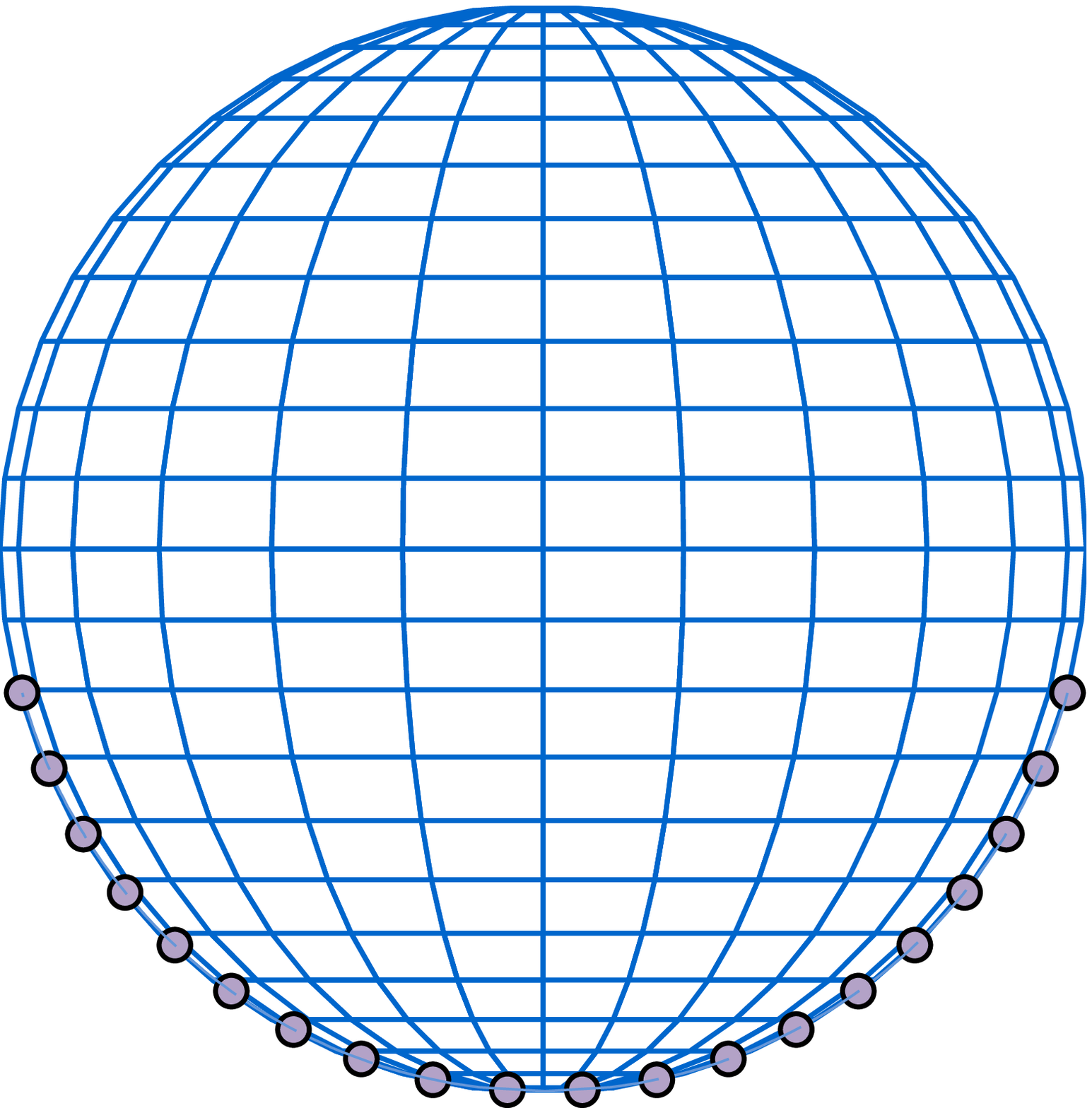}}
\subfigure[]{\includegraphics[width=0.22\columnwidth]{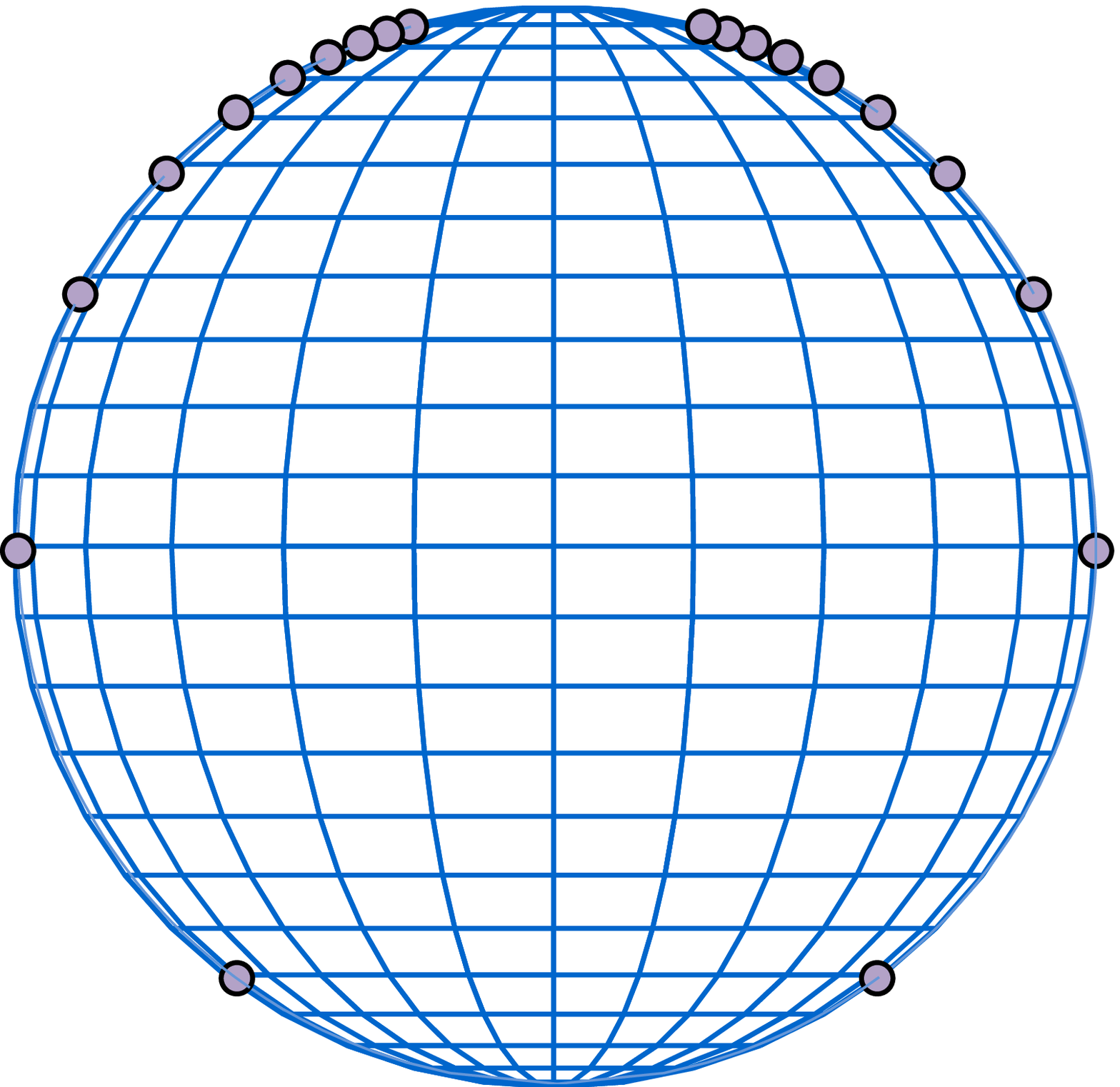}}
\caption{Bloch representation of CSA for the squeezed vacuum state with $N_c=20$ and (a) $\xi=0$, (b) $\xi=0.001$, (c) $\xi=0.01$, (d) $\xi=0.9$}
\label{CSSVS}
\end{figure}

\begin{figure}[b]
\centering
\subfigure[]{\includegraphics[width=0.24\columnwidth]{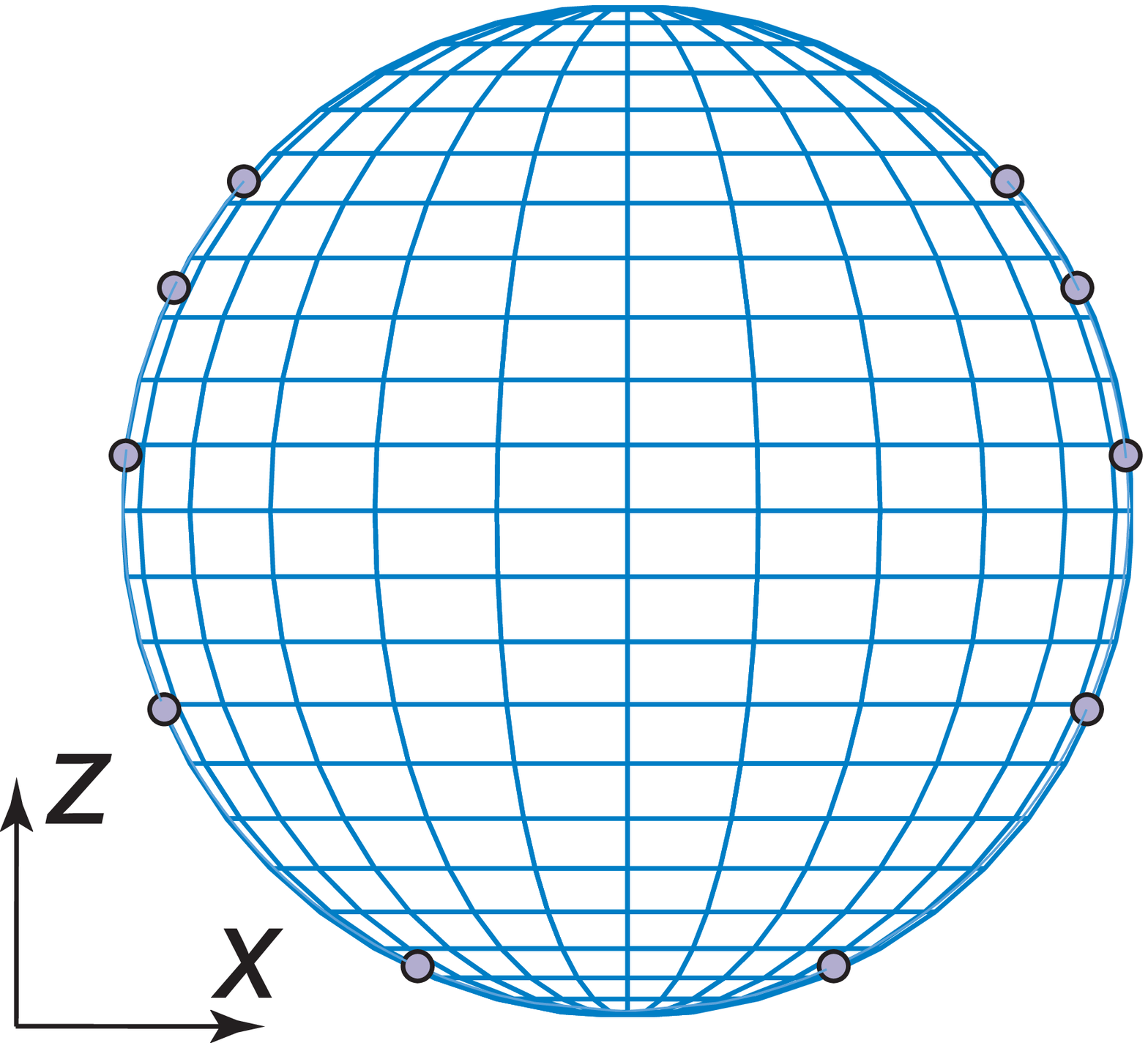}}
\subfigure[]{\includegraphics[width=0.22\columnwidth]{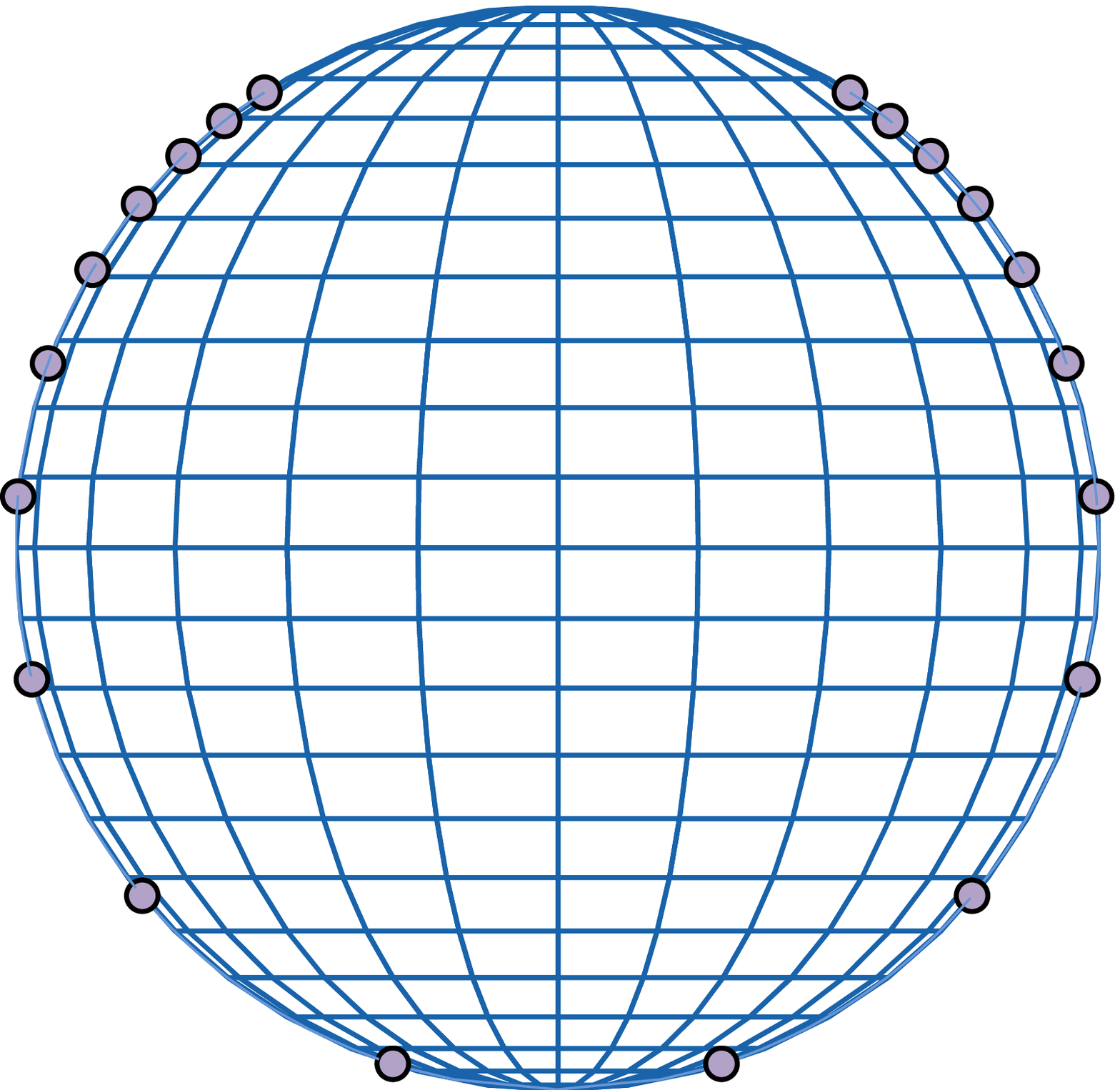}}
\subfigure[]{\includegraphics[width=0.22\columnwidth]{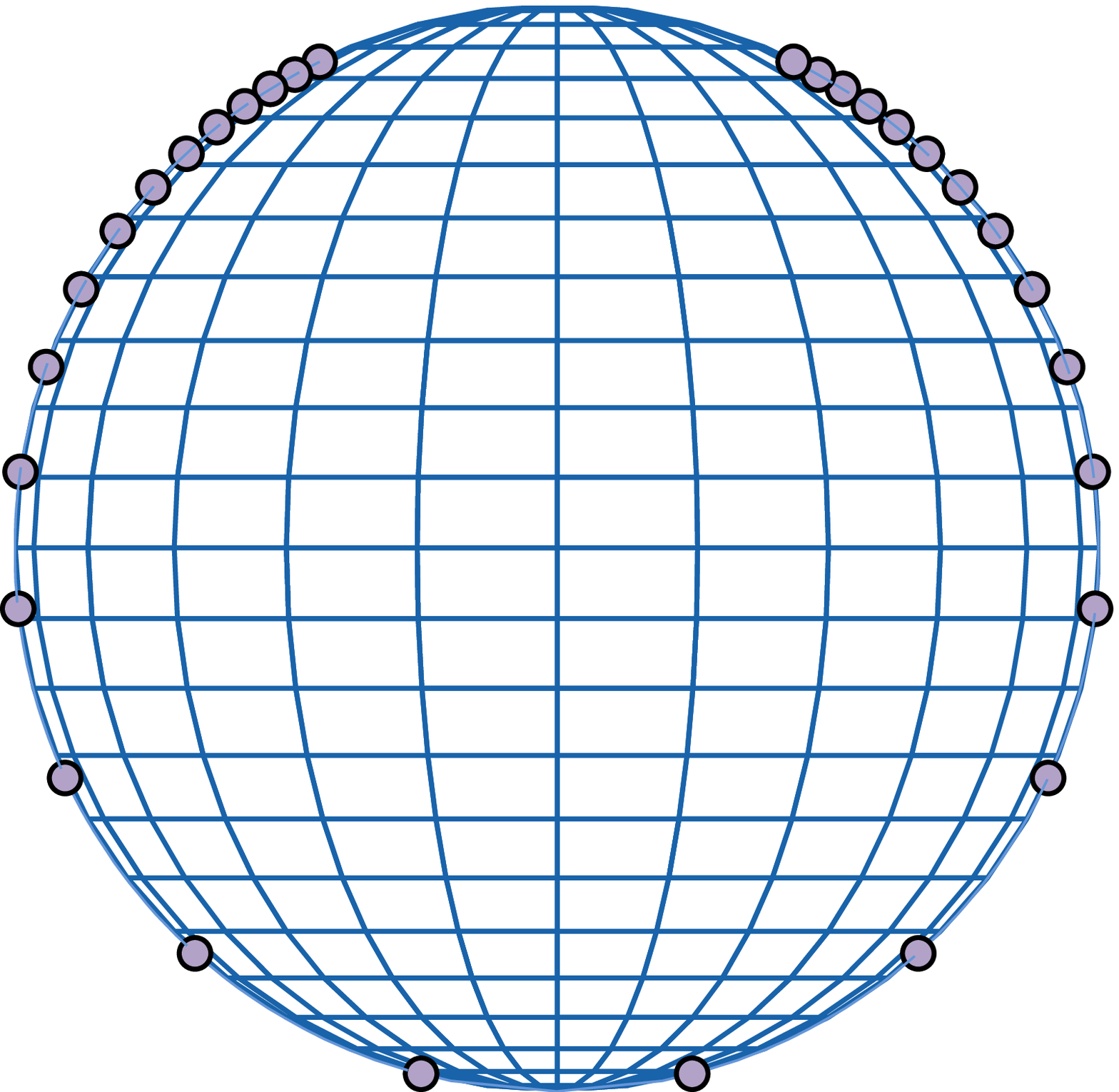}}
\subfigure[]{\includegraphics[width=0.22\columnwidth]{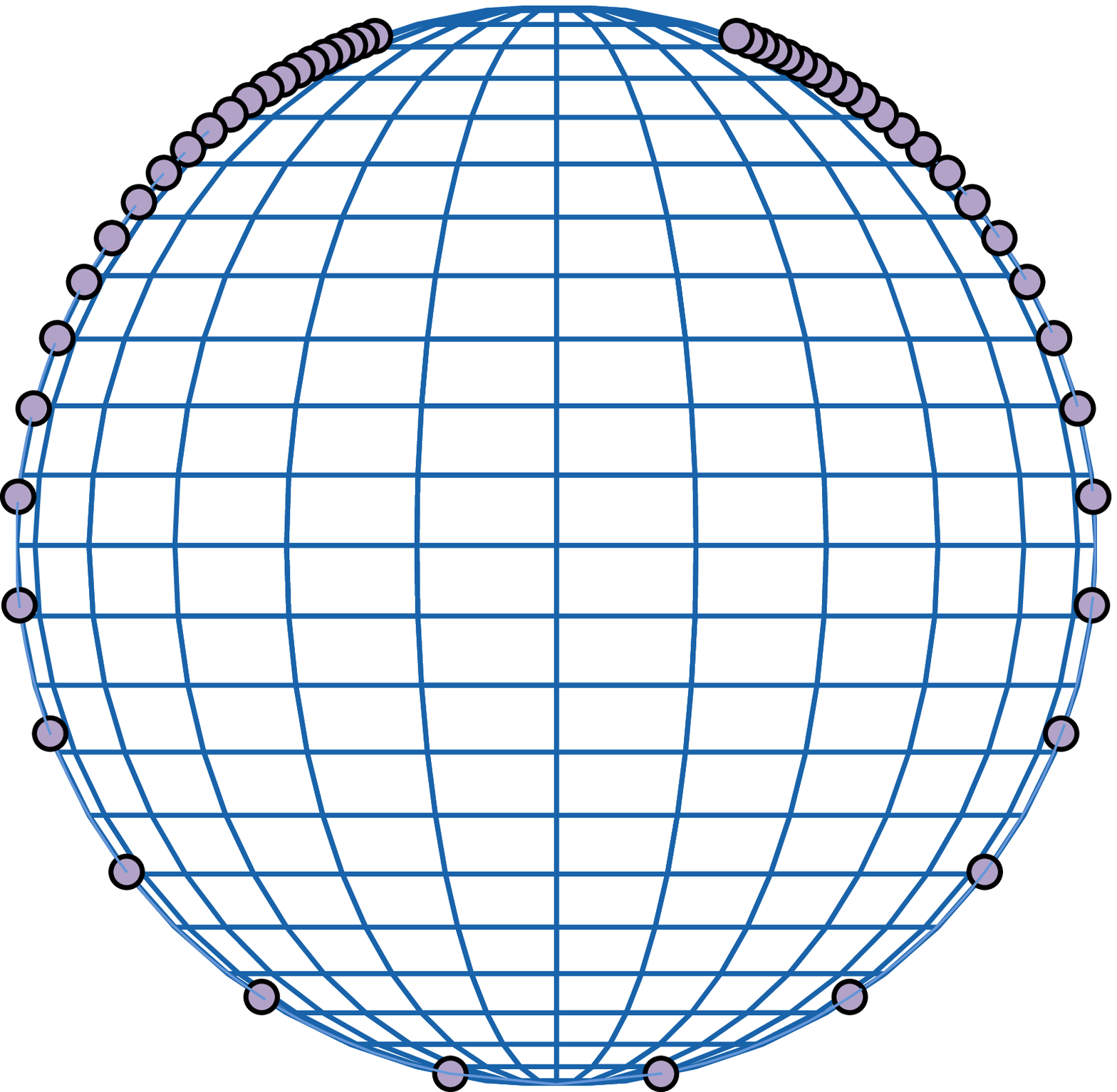}}
\caption{Bloch representation of CSA for the SMSV states with $\xi=0.2$ and different truncated numbers (a) $N_c=10$, (b) $N_c=20$, (c) $N_c=30$, and (d) $N_c=50$.}
\label{truncation}
\end{figure}

As the above three star equations becomes same for the states we have chosen, we here only consider bosonic squeezed vacuum state.
For even $N_c$, we have $N_c/2$ pairs of roots with phases $3\pi/2-\mathrm{Arg}(\xi)/2$ and $\pi/2-\mathrm{Arg}(\xi)/2$ and thus the stars are all distributed on a big circle of the Bloch sphere and  symmetric about $z$ axial, as shown in Fig. \ref{CSSVS}. As the squeezing parameter $|\xi|$ increases, the distribution of the stars varies from one overlapped points for the vacuum state $|0\rangle$ (see Fig. \ref{CSSVS}-(a)) to disperse points (see Fig. \ref{CSSVS}-(b) and (c)), and finally accumulate towards the north pole of the Bloch sphere (see Fig. \ref{CSSVS}-(d)). Thus, the increase of squeezing is intuitively represented by the moving of stars from south pole to north pole on the Bloch sphere.
Furthermore, we study  the influence of the truncation on the distribution of star.  We also show the CSA representation of the SMSV state with different truncation in Fig. \ref{truncation}. When the truncation is larger enough $(N_c\geq 20)$, the distributions of stars are similar as $N_c$ increases.

\begin{figure}[b]
\centering
\includegraphics[width=\columnwidth]{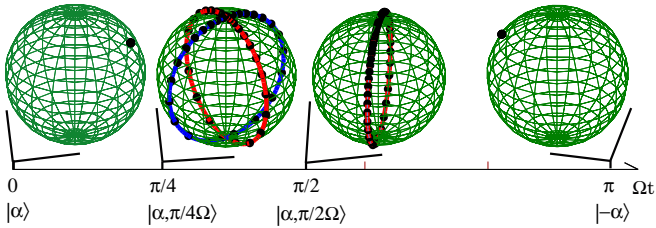}
\caption{CSA representation of CSA for the states $|\alpha,0\rangle$, $|\alpha,\pi/4\Omega\rangle$, $|\alpha,\pi/2\Omega\rangle$, $|\alpha,\pi/\Omega\rangle$with HW, SU(2) and SU(1,1) algebra, the parameters are chosen as $\alpha=2$, $n=50$}
\label{N2evolve}
\end{figure}

\section{Example: Quantum evolution in CSA representation}
To describe the quantum dynamics in CSA representation, we now consider a nonlinear Hamiltonian of the form\cite{Yurke1986}
\begin{equation}
 {H}=\omega  {\cal N}_F+\Omega  {\cal N}_F^2,
\end{equation}
where $\omega$ is the energy-level splitting for the linear part of the Hamiltonian, ${\cal N}_F^2$ is a nonlinear operator (such as $(a^\dag a)^2$ for HW which can be derived by a Kerr nonlinearity\cite{Gong2009}), and $\Omega$ is the strength of the nonlinear term. This model can be used to describe all of the three symmetries HW, SU(2) and SU(1,1) with $  {\cal N}_F={\cal N}=a^\dag  a$, $  {\cal N}_F=  {\cal N}_j=j+  J_Z$ and $  {\cal N}_F={\cal N}_k=-k+  K_0$, respectively. They all hold the relation $  {\cal N}|m\rangle=m|0\rangle$. Taking $\Omega  {\cal N}^2$ as the interaction part, the time evolution operator in the interaction picture takes the form $e^{-i\Omega t  {\cal N}^2}$. If we chose the initial state as the coherent state
\begin{equation}
|\alpha\rangle\sim\sum_{n=0}^{N_F}\frac{g_F(n)\alpha^n}{\sqrt{n!}}|n\rangle,
\end{equation}
where $|\alpha\rangle$  corresponds to $|\alpha_a\rangle, |\alpha_j\rangle, |\alpha_k\rangle$ for HW, SU(2) and SU(1,1), respectively, and the parameters are defined accordingly
\begin{equation}
\left\{
                      \begin{array}{ll}
                        g_a(n)=1 & N_c\rightarrow\infty,\\
                        g_J(n)=\sqrt{\frac{(2j)!}{(2j-2n)!}} & N_J=2j,\\
                       g_K(n)=\sqrt{\frac{\Gamma(2k+n)}{\Gamma(2k)}} & N_c\rightarrow\infty.\\
                      \end{array}
                    \right.\\
\label{example:paprameters}
\end{equation}
Therefore, the state in time $t$ can be written as
\begin{equation}
|\Psi(t)\rangle=e^{-i\Omega t  {\cal N}_F^2}|\alpha\rangle\sim\sum_{n=0}^{N_F}\frac{g_F(n)\alpha^ne^{-i\Omega n^2t}}{\sqrt{n!}}|n\rangle.
\label{tcs}
\end{equation}

Substituting Eq. (\ref{tcs}) into Eq. (\ref{su2star}), (\ref{HWstar}) and (\ref{su11star}) with the definition in Eq. (\ref{example:paprameters}), we have three equations of stars for HW, SU(2) and SU(1,1), respectively
\begin{equation}
\begin{aligned}
&\sum^{N_c}_{n=0}\frac{(-1)^n\alpha^ne^{-i\Omega n^2t}z^n}{(N_c-n)!n!}=0,\\
&\sum^{2j}_{n=0}\frac{(-1)^n\alpha^ne^{-i\Omega n^2t}x^n}{(2j-n)!n!}=0,\\
&\sum^{N_c}_{n=0}\frac{(-1)^n\alpha^ne^{-i\Omega n^2t}y^n}{(N_c-n)!n!}=0.
\label{example:equations}
\end{aligned}
\end{equation}
These three equations are identical when $N_C=2j$. Therefore, the distribution of the stars representing state $|\Psi(t)\rangle$ in CSA representation for the three different symmetries are same.

It is easy to find that this state is periodic with a period $2\pi/\Omega$ since $|\alpha,t+2\pi/\Omega\rangle=|\alpha,t\rangle$. In one period, the state $\Psi(t)\rangle$ is very interesting at some special time points and corresponds to some special distributions of stars by solving Eq. (\ref{example:equations}). At $t=0$, the state is initially on the coherent state $|\alpha\rangle$ with all of the stars overlap on one point (as shown on the sphere at the original point in Fig. \ref{N2evolve}); when the state evolves to time point $t=\pi/(4\Omega)$, the state becomes the superposition of four different coherent states as\cite{Yurke1986}
\begin{equation}
\left|\Psi\left(t=\frac{\pi}{4\Omega}\right)\right\rangle=\frac12[e^{-i\pi/4}|\alpha\rangle+|i\alpha\rangle-e^{-i\pi/4}|-\alpha\rangle+|-i\alpha\rangle].
\end{equation}

By Eq. (\ref{HWC}) and (\ref{HWstar}), the star equation for this superposition of four coherent states becomes
\begin{equation}
\begin{aligned}
&\sum_{n=0}^{[(N_c-1)/2]}e^{-i\pi/4}\begin{pmatrix}
                                                          N_c \\
                                                          2n+1 \\
                                                        \end{pmatrix}u^{2n+1}+\sum_{n=0}^{[N_c/2]}\begin{pmatrix}
                                                          N_c \\
                                                          2n \\
                                                        \end{pmatrix}(-1)^nu^{2n}\\
&=0
\end{aligned}
\end{equation}
with $u\equiv\alpha z$. When $N_c$ is very large, by numerical simulation, we find that the arguments of the roots $u_n$ of this equation can only take four phases $\pi/4$, $3\pi/4$, $5\pi/4$, $7\pi/4$. Therefore, all the stars will be distributed on two orthogonal circles as shown on the sphere at the point $\pi/4$ in Fig. \ref{N2evolve}.

As the state arrives at $t=\pi/(2\Omega)$, the state turns into a cat state\cite{Yurke1986}
\begin{equation}
\left|\Psi\left(t=\frac{\pi}{2\Omega}\right)\right\rangle=\frac {1}{\sqrt2}[e^{-i\pi/4}|\alpha\rangle+e^{i\pi/4}|-\alpha\rangle].
\end{equation}
Using Eq. (\ref{HWC}) and (\ref{HWstar}), the star equation for this superposition of two coherent states becomes
\begin{equation}
\begin{aligned}
e^{-i\pi/4}(1-\alpha z)^{N_c}+e^{i\pi/4}(1+\alpha z)^{N_c}=0.
\label{root4}
\end{aligned}
\end{equation}
Therefore, we have
\begin{equation}
\frac{(1+\alpha z)^{N_c}}{(1-\alpha z)^{N_c}}\equiv \lambda^{N_c}=e^{i\pi/2}.
\end{equation}
Define $\lambda=Ae^{i\phi}$, we have $A^{N_c}=1$ and $e^{iN_c\phi}=e^{i\pi/2}$, which implies $A=1$ and $\phi=(2n\pi+\pi/2)/N_c$ with $n=0,1,\cdots,N_c-1$.
Thus, the $n$-th root of $z$ satisfies the relation
\begin{equation}
\frac{1+\alpha z_n}{1-\alpha z_n}=\exp\left[\frac{i(4n+1)\pi}{2N_c}\right].
\end{equation}
Therefore, the roots of Eq. (\ref{root4}) can be derived as
\begin{equation}
z_n=\frac1\alpha\frac{e^{i(4n+1)\pi/2}-1}{e^{i(4n+1)\pi/2}+1}=\frac i\alpha\tan\left[\frac{(4n+1)\pi}{4N_c}\right]
\end{equation}
By the definition $z_n=\tan\frac{\theta_n}{2}e^{i\phi_n}$, the spherical coordinates of the stars can be given by
\begin{equation}
\begin{aligned}
\theta_n&=2\arctan\left(\frac1\alpha\left|\tan\left[\frac{(4n+1)\pi}{4N_c}\right]\right|\right)\\
\phi_n&=\left\{
          \begin{array}{cc}
          \pi/2-\arg(\alpha) & n\leq[(N_c-1)/2] \\
            \phi_n=3\pi/2-\arg(\alpha) & n>[(N_c-1)/2] \\
          \end{array}
        \right.
\end{aligned}
\end{equation}
The numerical results are  shown on the sphere at the point $\pi/2$ in Fig. \ref{N2evolve}), and the stars are distributed
on one large circle.

When the state evolves a half period, the state reverts to a single coherent state as
\begin{equation}
\left|\Psi\left(t=\frac{\pi}{\Omega}\right)\right\rangle=|-\alpha\rangle
 \end{equation}
with $n$ coincided stars on the Bloch sphere (as shown on the sphere at the point $\pi$ in Fig. \ref{N2evolve}). Thus, the period evolution of a quantum state can be perfectly reflected by the period changes of stars on the Bloch sphere. Moreover, if $2\pi$ is dividable by $\Omega t$, the stars for the state at this moment are distributed on several circles and the state can be written as the superposition of several coherent states. Furthermore, according to Eq. (\ref{example:equations}), these interesting phenomena can be observed in all of these three different symmetries.

\section{Conclusions}
The Majorana representation provide us a geometric tool to study the quantum states with SU(2) symmetry and their evolutions. Our study here is to show how can we extend this elegant method to the system with both finite and infinite dimensions. We found that the key of the answer is the coherent state. The definitions of coherent states in different kind of symmetries inspired us a method to build the representation by ladder and number operators and provide a reference state to our representation. By study three different symmetries, we show this coherent-state approach of Majorana representation can well characterize squeezed states for different symmetries and the dynamical evolution of a quantum state. Furthermore, there are more symmetries like SU(N) $(N>2)$ need to be further studied with the CSA representation and the coherent-state approach method will be very helpful for further studies.

\ \ \\
This work is supported by the National Fundamental Research Program of China (Contact Nos. 2013CBA01502), the National Natural Science Foundation of China (Contact Nos. 11374040, 11575027,11475146, and 11405008)£¬ and the Fundamental Research Funds for the Central Universities (Grants No. 2412015KJ009)

\end{document}